\documentstyle[11pt]{article}

\title{Chiral Technicolor and Precision Electroweak Measurements}
\author{ John Terning\\ Department of Physics, Boston University\\ 590
Commonwealth Ave., Boston MA  02215}
\date{October 6,1994}
\begin{document}
\setlength{\baselineskip}{24pt}
\maketitle
\begin{picture}(0,0)(0,0) \put(295,270){BUHEP-94-27} \end{picture}
\vspace{-12pt}

\begin{abstract}I consider the possibility that electroweak  symmetry is
broken by a strongly interacting chiral gauge theory. I argue that some of
the discrepancies between precision electroweak measurements and the
predictions of QCD-like technicolor models can be resolved if technicolor
is a chiral gauge theory.  I present a toy technicolor model which
demonstrates this idea, and gives  $ m_t \gg m_b$, with a small value for
$\Delta\rho_* \equiv \alpha T $,
and small corrections to $Z \rightarrow b \overline{b}$. \end{abstract}

 \section{Introduction} The origin of electroweak symmetry breaking remains
 a basic problem of particle
physics. There has recently been a small revival of
interest in the technicolor (TC) approach to electroweak symmetry
breaking \cite{TC}, and in building realistic extended technicolor (ETC)
models which reproduce the quark-lepton mass spectrum without
phenomenological disasters like flavor changing neutral currents
\cite{ETCmodels,AppelTern}.  Effort in this direction has focused on avoiding
the
pitfalls of TC  models which relied on naively scaling-up the
properties of QCD.  This work has included examining theories possessing:
smaller $\beta$ functions (a.k.a. walking \cite{walking}), scalars
\cite{scalars}, near-critical
four-fermion interactions (strong ETC \cite{strongETC}), multiple scales
of electroweak symmetry breaking (multiscale models \cite{multi}),and
GIM symmetries (Techni-GIM \cite{TechniGIM}).  Although there has been
some measure of success along these lines, precision electroweak
measurements continue to put tighter constraints on TC models, and if
experiments converge near the current central values for $S$, $T$, $m_t$,
and the partial width  $\Gamma(Z\rightarrow b \overline{b})$, then life
will be extremely difficult for model builders.  The problems of
compatibility with precision electroweak measurements, combined with
the general awkwardness of obtaining a reasonable fermion spectrum
seem to suggest that some new ingredient is required for a truly realistic
ETC model.

In all the variants of TC mentioned above, there is one important common
assumption: that TC is vector-like.  That is (like QCD) left-handed
and right-handed technifermions have identical TC gauge interactions.  In
this paper I will start to explore what can happen when this
assumption is dropped.  In the next section I describe in more detail the
problems of vector-like TC models, and suggest how these problems could
be ameliorated in chiral TC models. In the third and fourth sections I examine
in some
detail a toy chiral TC model which demonstrates some of the general features
of this approach.

\section {Problems with Vector-Like TC}
There are three potential conflicts between the current crop of TC
models and experiment that I would like to focus on:
\newline $\bullet$ current measurements of the ratio of
$\Gamma(Z\rightarrow b \overline{b}) $ to $\Gamma(Z\rightarrow q
\overline{q}) $ are 2\% above the standard model prediction (with an error
around 1\%), while simple TC models  suggest that this ratio should be
several percent below the standard model \cite{Zbb},
\newline $\bullet$ the $S$ parameter seems to be negative, or at least small,
whereas simple QCD-like TC
models prefer  positive values \cite{PT} of order one,
\newline $\bullet $ the $t$ quark is very heavy, but in a TC framework it
is difficult to make $ m_t \gg m_b $, and it is even more difficult to do
this while keeping $ \Delta\rho_* \equiv \alpha T$ small
\cite{Sikivie,isospin}.

Each of these potential problems can be individually circumvented. Models
where $SU(2)_L$ does not commute with ETC interactions can
produce the opposite sign correction to  the
ratio of $\Gamma(Z\rightarrow b \overline{b}) / \Gamma(Z\rightarrow q
\overline{q}) $ \cite{posZbb}.
Alternatively, the correction can be reduced by fine-tuning ETC interactions
\cite{Evans}.  There are
several schemes which can produce negative contributions to $S$
\cite{negS,revenge}.  One can arrange a model to make $m_t \gg m_b$  by
fine-tuning ETC interactions \cite{strongETC} or by having different ETC scales
\cite{EN}
for the $t_R$ and the $b_R$.  However, these three problems are all
intimately related to isospin breaking, and it is possible that there is a
common solution to all three problems.  It is this possibility that leads to
the consideration of chiral TC models. In order to see why, it will be
useful to consider the three problems in more detail

One possible solution to the $Z\rightarrow b \overline{b}$ problem is that
the ETC corrections are actually much smaller than is expected in
vector-like TC models.  Recall that the size of the correction is
proportional to
\begin{equation}
g_{ETC}^2 {{f^2}\over{M_{ETC}^2}} \approx {{m_t}\over{4
\pi f}} ~,
\label{Zbbmt} \end{equation}
 (where $g_{ETC}$ is the ETC gauge coupling,
$M_{ETC}$ is the mass of the ETC gauge boson, and $f$ is the technipion
decay constant) which comes from assuming that the technifermions
which couple to the $b$ quark are approximately degenerate with those
that couple to the $t$ quark \cite{Zbb}. The implicit assumption being that if
this were
not the case $ \Delta \rho_* \equiv \alpha T$ would be large. With this
assumption the size of the correction is related to the $t$ quark mass.
Given that $m_t$ is probably near 175 GeV, and $f < 246$ GeV,  the
correction is quite large.

If, however, the up-type and down-type technifermions were not
degenerate, then there would be two different  technipion decay constants:
$f_U$ and $f_D$.  Then equation (\ref{Zbbmt}) would be replaced by:
\begin{equation}
g_{ETC}^2 {{f_D^2}\over{M_{ETC}^2}}  ~.
\label{Zbbmb} \end{equation}
If the ETC scales are the same for the $t$ and the $b$, then this suggests a
smaller correction than is usually expected. Of course it remains to
explain why $T$ is small in such a model.

Next, consider the problem of negative $S$.  A simple mechanism for
producing negative contributions to $S$ in a one-family model was
suggested in ref. \cite{revenge}.  There a large mass splitting between the
technielectron and the technineutrino gave a negative contribution to $S$.
In order to keep $T$ small, the technileptons had to be much lighter than
the techniquarks, and the techniquarks had to be roughly degenerate.
Since, naively, the contribution to the $W^\pm$ and $Z$ masses (and hence to
$T$) scales as the technifermion mass squared  (or technipion decay
constant squared) the degenerate techniquarks provided the bulk of the
gauge boson masses, and kept $T$ small.  Note however that there are
three times as many techniquarks as technileptons, and the techniquarks
(not to mention the
pseudo-Nambu-Goldstone bosons) give a positive contribution to $S$, so that
although $S$ could be reduced from naive expectations, it could not be
made substantially negative.  Now, if the up-type techniquark was much
heavier than the down-type techniquark (as was suggested by the
consideration of the $Z\rightarrow b \overline{b}$ problem), then the
techniquark contribution to $S$ would be reduced.  Again, this solution, on
the face of it, seems incompatible with a small value for $T$.

Finally, consider the problem of the large $t$ quark mass.  Obviously the
simplest way to make $m_t \gg m_b$ is to (again) have an up-type
techniquark that is much heavier than the down-type techniquark.  Thus all
three problems are at least partially alleviated by isospin breaking in the
techniquark sector.
The price of this solution seems to be an even bigger problem: the wrong
masses for the $W^\pm$ and the $Z$, or in other words a very large value
for $T$.

Traditional TC models have always done a beautiful job of getting
the correct masses for the $W^\pm$ and $Z$.  The problems with TC
arise when one tries to use these same technifermions to give masses to
the quarks and leptons. Once we assume
that TC is vector-like, the $W^\pm$ and $Z$
masses are guaranteed to be in the
correct ratio by the unbroken $SU(2)_V$ custodial isospin group.
When considering the problems of precision
electroweak measurements above, I was lead to consider a spectrum of
technifermions which is compatible with quark and lepton masses, but
gives the wrong gauge boson masses.  A resolution of this discrepancy is
obvious, if  perhaps inelegant.  If there are two types of technifermions (that
is
technifermions in two different representations of the TC gauge group)
then we can imagine that one type gives masses to the gauge
bosons, while the other type primarily gives masses to the quarks and
leptons. (This is what happens in multiscale TC models \cite{multi}.)
Since both types of technifermions must carry $SU(2)_L$
quantum numbers, they will in fact both contribute to the gauge boson
masses.  However it is the heavier
technifermions that will provide the bulk of the gauge boson masses, and
it is these technifermions that must have a good custodial isospin
symmetry.  Ensuring that only the light (isospin breaking) sector
contributes to quark and lepton masses depends entirely on how the TC
gauge group is embedded in the ETC gauge group.

The remaining problem is how to break isospin symmetries in the light
technifermion sector.  One could imagine putting the right-handed up-type
and down-type technifermions in different ETC representations which
decompose into equivalent TC representations \cite{AppelTern,multi,Sikivie,EL}
(e.g. when $SU(3)_{ETC}$ breaks to $SU(2)_{TC}$, a ${\bf 3}$ and
${\bf \overline{3}}$
both decompose to ${\bf 1} + {\bf 2}$). Then, if the ETC interactions are
strong enough, they can produce a substantial  splitting of technifermion
masses. A more direct approach is to put the right-handed up-type and
down-type technifermions in different TC representations, which implies
that TC is a chiral gauge theory.
In the next section I will present a toy model which exemplifies these
ideas.

\section{A Toy Model}
The model is anomaly free and asymptotically free, so
that although it is not complete (for example, there are numerous massless
Nambu-Goldstone bosons), it is at least internally
consistent.  The gauge group of the model is
$SU(5)_{ETC} \otimes
SU(3)_{C}\otimes SU(2)_L \otimes U(1)_Y$, with the fermion content
taken to be\footnote{I will use
charge-conjugates of the right-handed fields.}:
\begin{eqnarray}
\begin{array}{ccc} ({\bf 5},{\bf 3},{\bf 2})_{1 \over 3} & ({\bf
\overline{5}},{\bf \overline{3}},{\bf 1})_{-{4 \over 3}} & ({\bf 5},{\bf
\overline{3}},{\bf 1})_{2 \over 3}
\\  \\
({\bf \overline{5}},{\bf 1},{\bf 2})_{-1} & ({\bf 10},{\bf 1},{\bf 1})_0
& ({\bf \overline{5}},{\bf 1},{\bf 1})_{2}
\end{array}
\label{ferm1}
\end{eqnarray}
\begin{equation} \begin{array}{ccc}
({\bf 10},{\bf 1},{\bf 2})_0 & ({\bf 10},{\bf 1},{\bf 1})_{-1} & ({\bf 10},{\bf
1},{\bf 1})_{1}
\end{array}\label{ferm2}
\end{equation}
\begin{equation} \begin{array}{cc} ({\bf \overline{15}},{\bf 1},{\bf 1})_0
&  ({\bf 5},{\bf 1},{\bf 1})_0  ~. \end{array} \label{ferm3}
\end{equation}
The ${\bf 10}$ is the familiar antisymmetric tensor representation of
$SU(5)_{ETC}$, while the
${\bf 15}$ is the symmetric tensor representation.
With the fermions listed above, the $SU(5)_{ETC}$ gauge anomaly cancels
since (in a certain normalization) the ${\bf 5}$'s and ${\bf 10}$'s each
contribute $+1$ to the anomaly while the ${\bf \overline{15}}$
contributes\footnote{For $SU(N)$, $N \ge 3$, the antisymmetric tensor
contributes $N-4$ to the anomaly,
while the symmetric tensor contributes $N+4$.}
$-9$.  The third generation of quarks and leptons, and their
associated technifermions, are contained in the representations listed in
(\ref{ferm1}). The fermions listed in (\ref{ferm2}) will turn out to
contain the heavy technifermion sector.

This model is an example of a
chiral gauge theory, and I expect that when the  $SU(5)_{ETC}$ gauge
interaction becomes strong it will break itself by forming a fermion
condensate that is a gauge-non-singlet
(this phenomena is referred to as ``tumbling"
\cite{MAC1} in the literature). Folklore suggests that the condensate will
form in the most attractive channel (MAC) \cite{MAC1,MAC2}, as
determined by the examining the relative attractive strength of the exchange of
one
massless gauge boson in the various channels. In this approximation the
strength of the channel ${\bf R_1} \times {\bf R_2} \rightarrow {\bf R}$
is proportional to
\begin{equation}
\Delta C_2({\bf R_1} \times {\bf R_2} \rightarrow {\bf R})
 \equiv C_2({\bf R_1}) + C_2({\bf R_2}) - C_2({\bf R})~,
\label{DeltaC2}
\end{equation} where $C_2({\bf R})$ is the quadratic Casimir of
representation ${\bf R}$.  The MAC in this model is ${\bf \overline{15}}
\times {\bf 5} \rightarrow {\bf \overline{5}}$ with $\Delta C_2 = 28/5$.
The next most attractive channels are ${\bf 5} \times {\bf \overline{5}}
\rightarrow {\bf 1}$ and ${\bf 10} \times {\bf 10} \rightarrow {\bf
\overline{5}}$, both with $\Delta C_2 = 24/5$.  I will assume that
condensation occurs in the MAC, which breaks $SU(5)_{ETC}$ down to
$SU(4)_{TC}$.

Below the ETC scale (noting  that under
$SU(4)_{TC}$ the ${\bf\overline{15}}$ decomposes as ${\bf 1} + {\bf
\overline{4}}
+ {\bf \overline{10}}$) we have the following fermions (labeled by
$SU(4)_{TC} \otimes SU(3)_{C} \otimes SU(2)_L \otimes U(1)_Y$) :
\begin{eqnarray}
\begin{array}{ccc}
({\bf 1},{\bf 3},{\bf 2})_{1 \over 3} &
({\bf 1},{\bf \overline{3}},{\bf 1})_{-{4 \over 3}} & ({\bf 1},{\bf
\overline{3}},{\bf 1})_{2 \over 3}  \\
 (t,b)_L  & t_R^c
 & b_R^c \\   \\
 ({\bf 4},{\bf 3},{\bf 2})_{1
\over 3} & ({\bf \overline{4}},{\bf \overline{3}},{\bf 1})_{-{4 \over 3}} &
({\bf 4},{\bf \overline{3}},{\bf 1})_{2 \over 3}  \\
 (U,D)_L  & U_R^c  & D_R^c  \\
\end{array}
\label{ferm421Q}
\end{eqnarray}
\begin{eqnarray}
\begin{array}{ccc} ({\bf 1},{\bf 1},{\bf 2})_{-1} &
({\bf 6},{\bf 1},{\bf 1})_0 & ({\bf 1},{\bf 1},{\bf 1})_{2}
 \\
(\tau,\nu_\tau)_L  &
  &
\tau_R^c \\ \\
 ({\bf \overline{4}},{\bf 1},{\bf 2})_{-1} & ({\bf 4},{\bf 1},{\bf 1})_0
& ({\bf \overline{4}},{\bf 1},{\bf 1})_{2}   \\
(N,E)_L &  N_R^c & E_R^c \\
\end{array}
\label{ferm421L}
\end{eqnarray}
\begin{eqnarray}
\begin{array}{ccc} ({\bf 4},{\bf 1},{\bf 2})_0 & ({\bf 4},{\bf 1},{\bf 1})_{-1}
& ({\bf 4},{\bf 1},{\bf 1})_{1} \\
({\bf 6},{\bf 1},{\bf 2})_0 & ({\bf 6},{\bf 1},{\bf 1})_{-1} & ({\bf 6},{\bf
1},{\bf 1})_{1}
\end{array}
\label{ferm422}
\end{eqnarray}
\begin{equation} \begin{array}{c}
({\bf  \overline{10}},{\bf 1},{\bf 1})_0    ~.
\end{array} \label{ferm423} \end{equation}
Note that $SU(4)_{TC}$ is an asymptotically free chiral gauge theory.
As we descend the energy scale, further fermion condensation
should occur.  I expect that the next condensate will form in the new MAC,
which is ${\bf 6} \times {\bf 6} \rightarrow {\bf 1}$, with $\Delta C_2 =
5$.  This breaks $SU(2)_L \otimes U(1)_Y$ down to $U(1)_{em}$, and makes some
extra technineutrinos in line (\ref{ferm421L}) heavy.  I will refer to the
technifermions
which condense at this scale as the heavy technifermions.  Below
the electroweak symmetry breaking scale we have the following fermions
(labeled according to $SU(4)_{TC} \otimes SU(3)_C \otimes U(1)_{em}$):
\begin{eqnarray}
\begin{array}{cccc} ({\bf 1},{\bf 3})_{2 \over 3} & ({\bf 1},{\bf 3})_{-{1
\over 3}} & ({\bf 1},{\bf \overline{3}})_{-{2 \over 3}} & ({\bf 1},{\bf
\overline{3}})_{1 \over 3}
 \\
 t_L  & b_L & t_R^c & b_R^c \\  \\
 ({\bf 4},{\bf 3})_{2 \over 3} & ({\bf 4},{\bf 3})_{-{1
\over 3}} & ({\bf \overline{4}},{\bf \overline{3}})_{-{2 \over 3}} & ({\bf
4},{\bf \overline{3}})_{1 \over 3}  \\
U_L & D_L  & U_R^c & D_R^c \\
 \\ \end{array}
\label{ferm41Q}
\end{eqnarray}
\begin{eqnarray}
\begin{array}{cccc} ({\bf 1},{\bf 1})_{0} & ({\bf
1},{\bf 1})_{-1} &  & ({\bf 1},{\bf 1})_{1}
 \\
 \tau_L & \nu_{\tau L}&   & \tau_R^c \\ \\
({\bf \overline{4}},{\bf 1})_{0} & ({\bf \overline{4}},{\bf 1})_{-1} & ({\bf
4},{\bf 1})_0 & ({\bf \overline{4}},{\bf 1})_{1}  \\
N_L  & E_L
& N_R^c  & E_R^c \\
\end{array}
\label{ferm41L}
\end{eqnarray}
\begin{eqnarray}
\begin{array}{cccc}
({\bf 4},{\bf 1})_{{1 \over 2}} & ({\bf 4},{\bf 1})_{-{1 \over 2}} & ({\bf
4},{\bf 1})_{-{1 \over 2}} & ({\bf 4},{\bf 1})_{{1 \over 2}}
\end{array}
\label{ferm42}
\end{eqnarray}
\begin{equation}
\begin{array}{c}  ({\bf \overline{10}},{\bf 1})_0
  ~.
\end{array}
\label{ferm43}
\end{equation}

Even though the electroweak gauge symmetry has been broken, the
effective $SU(4)_{TC}$ gauge theory is chiral, and further condensates are
to be expected.   The MAC is now ${\bf \overline{10}} \times {\bf 4}
\rightarrow {\bf \overline{4}}$ with $\Delta C_2 = 9/2$. This condensate
would break $SU(4)_{TC}$ down to $SU(3)$. However, here I will assume
that the MAC analysis is not correct.   I imagine that  the next condensation
occurs in the channel ${\bf 4} \times {\bf 4} \rightarrow {\bf 6}$, which
breaks $SU(4)_{TC}$ down to $Sp(4)_{RTC}$.  (See the Appendix for
a justification of this assumption.) I will refer to this unbroken
subgroup as the residual technicolor (RTC) group. With this pattern of
condensation, all the light technifermions
associated with the third family will become
massive. Below the TC breaking scale we have (labeled according to
$Sp(4)_{RTC} \otimes SU(3)_C  \otimes U(1)_{em}$) the following fermion
content:
\begin{eqnarray}
\begin{array}{cccc} ({\bf 1},{\bf 3})_{2 \over 3} & ({\bf 1},{\bf 3})_{-{1
\over 3}} & ({\bf 1},{\bf \overline{3}})_{-{2 \over 3}} & ({\bf 1},{\bf
\overline{3}})_{1 \over 3}  \\
 t_L & b_L & t_R^c & b_R^c  \\  \\
({\bf 1},{\bf 1})_{0} & ({\bf 1},{\bf 1})_{-1} &   & ({\bf 1},{\bf 1})_{1}
 \\
 \tau_L & \nu_{\tau L}
&  &
\tau_R^c \\ \end{array} \label{fermSp41}
\end{eqnarray} \begin{equation} \begin{array}{c}
({\bf 10},{\bf 1})_0  ~. \end{array} \label{fermSp43}
\end{equation}

Finally, at this stage there is nothing to prevent condensation in the
channel ${\bf 10} \times {\bf 10} \rightarrow {\bf 1}$, and we are left
with only the usual third family.  We note that because
of the chiral structure, the only third generation fermion that receives a mass
is the $t$ quark.  This is, in fact, a good first approximation to
the observed spectrum.  To give masses to the other fermions, one would have to
embed
the $SU(5)_{ETC}$ gauge group in a larger group, but I will not pursue this
possibility,
since this is only a toy model.  In the next section I will attempt to analyze
the technifermion
spectrum in some detail, not because this model is especially interesting as it
stands, but in order to
illustrate some generic features of chiral TC models.

\section{The Technifermion Spectrum}

Given the current state of field theory technology, we will have to be
satisfied with the traditional analysis of the Schwinger-Dyson  equation for
the self-energy (in the traditional ladder approximation, for a discussion of
the reliability of this approximation see ref. \cite{ALM}).     In the unbroken
theory the
self-energy graph of a fermion  in representation
${\bf R}$ is proportional to $C_2({\bf R})$, which comes
from summing over the square of gauge generators.   In chiral gauge theories
the left-handed and right-handed-conjugate
fermions can be in different representations (${\bf R_1}$ and $ {\bf R_2}$
say), and the
condensate is not necessarily a singlet (call it  ${\bf R}$).  Neglecting the
gauge boson masses for a
moment, the factor $C_2({\bf R})$ must be replaced by
${1 \over 2}  \Delta C_2({\bf R_1} \times {\bf R_2} \rightarrow {\bf R})$
as given in equation (\ref{DeltaC2}). In a broken gauge theory (that is when
the representation, ${\bf R} $,
of the condensate is not a gauge singlet) we must separate the gauge bosons
into their representations under
the unbroken subgroup, since in general these different representations will
have different masses. For the
toy model I am discussing, the gauge bosons of $SU(5)_{ETC}$ decompose under
$Sp(4)_{RTC} $ as
${\bf 10} + {\bf 5} + {\bf 4}
+{\bf 4} + {\bf 1}$, where the  ${\bf 10}$ are the massless gauge bosons of
$Sp(4)_{RTC} $, the ${\bf
4}$'s and ${\bf 1}$ are the heavy ETC gauge bosons, and the ${\bf 5}$ are the
broken TC gauge bosons.
Thus I will write
\begin{equation}
\Delta C_2({\bf R_1} \times {\bf R_2} \rightarrow {\bf R})
= \sum_{\bf G} \Delta C_2^{\bf G} ~,
\end{equation}
where the gauge boson representation, ${\bf G}$, runs over $ {\bf 1}$,$ {\bf
4}$,${\bf 5} $, and ${\bf
10}$.
 For the sake of brevity the values for $\Delta C_2^{\bf G}$ in the various
channels are presented in Table
1.

\begin{table}[htbp]
\begin{center}
\begin{tabular}{|c||c||c||c||c||c||c|}\hline\hline
fermion & channel &\strut $\Delta C_2^{\bf 10}$ & $\Delta C_2^{\bf 5}$ &
$\Delta C_2^{\bf 4}$  & $\Delta
C_2^{\bf 1}$ &$\Delta C_2$ \\
\hline\hline
$U$  &${\bf 5} \times {\bf \overline{5}} \rightarrow {\bf 1}$&
$\vphantom{\Sigma}$ ${5\over 2}$& ${5\over 4}$ & $1$&
${1\over 20}$ &${24\over 5}$ \\ \hline
$t$   &${\bf 5} \times {\bf \overline{5}} \rightarrow {\bf 1}$&$0$&$0$ & $4$
&${4\over 5}$
&${24\over 5}$\\ \hline
$D$  &${\bf 5} \times {\bf 5} \rightarrow {\bf 10}$&${5\over 2}$& $-{5\over 4}$
& $0$ &
 ${1\over 20}$&${6\over 5}$  \\ \hline
$b$   &${\bf 5} \times {\bf 5} \rightarrow {\bf 15}$& $0$ &$0$ & $0$ &
$-{4\over 5}$ &$-{4\over 5}$
\\ \hline
$N$ &$ {\bf \overline{5}} \times {\bf 10}\rightarrow {\bf 5}$& ${5\over 2}$&
${5\over 4}$ & $0$& $-
{3\over 20}$ &${18\over 5}$ \\ \hline
$E$  &${\bf \overline{5}} \times {\bf \overline{5}} \rightarrow {\bf
\overline{10}}$&${5\over 2}$& $-
{5\over 4}$ & $0$ &
 ${1\over 20}$&${6\over 5}$  \\ \hline
$\tau$   &${\bf \overline{5}} \times {\bf \overline{5} }\rightarrow {\bf
\overline{15}}$& $0$ &$0$ &
$0$ & $-{4\over 5}$ &$-{4\over 5}$  \\
\hline\hline
\end{tabular}
\end{center}
\caption{Self-energy coefficients for the various fermions.  The condensation
channels are classified in
terms of $SU(5)_{ETC}$ representations.  The $\nu_\tau$ is not listed since
there no right-handed
$\nu_\tau$.}
\label{table}
\end{table}
It is worth pointing out several features in Table  1.  First note that masses
for the $D$ and the $b$
correspond to  different condensation channels.  This is a generic feature of
chiral TC models.  The $b$
remains  massless in this model, since there is no condensate which transforms
as a ${\bf 15}$ (a similar
argument applies for the $\tau$).

Secondly, note that as far as the massless RTC gauge bosons (in the ${\bf 10}$)
are concerned, all the light technifermions are
identical, so the critical coupling for this interaction is the same for all
technifermions.  However the broken
TC gauge bosons have different couplings to different technifermions.  Consider
the $U$ and $D$
techniquarks at energies above the TC breaking scale. If the TC gauge coupling
is close to critical for the
$U$, it will be far below critical for the $D$.  The repulsive interaction due
to the broken TC gauge bosons
will cause the self-energy of the $D$ to fall much more rapidly than that of
the $U$, which experiences
attractive interactions, and hence the self-energy at the ETC scale will be
larger for the $U$ than the $D$.
This effect will be enhanced in  chiral TC models that enjoy walking.  This is
an important
generic effect in chiral TC models since it reduces the amount of isospin
breaking in techniquark masses
that is needed (in a class of models) to explain the $t$-$b$ mass splitting.
The reasoning goes as follows.
If the $t$ and $b$ have the same ETC scale, then scaling from QCD gives:
\begin{equation}
{f_D\over f_U} = \left( {m_b\over m_t}\right)^{{1\over 3}} ~.
\label{ratio}
\end{equation}
This relation will {\em not} hold in chiral TC models, since the self-energies
fall off at different rates.  (Note that this ratio becomes even smaller in
vector-like, walking models.)
Since $m_t$ and $m_b$ are related to $U$ and $D$ self energies at the ETC
scale, and the $D$ self-energy
falls faster, the $b$ can be much lighter than the $t$, even though $f_D$ and
$f_U$ are fairly close
together.

Finally note that the $U$ and $t$ have especially strong ETC interactions
(corresponding to  $\Delta
C_2^{\bf 4}$ in Table 1), which is useful for enhancing the $t$ mass.  This
feature seems to be more model
dependent.  Given that $U$ has more attractive TC interactions and strong ETC
interactions, I expect that
the $U$ can be substantially heavier than the $D$.  In a model that could
produce a reasonable $t$ mass,
one would naively expect the difference between the $U$ and $D$ masses to be of
order $m_t$.  Such a
splitting is a severe problem in QCD-like TC models, because of the large
contribution to $T$.  Here, and
in multiscale models in general, things are not as bad as in QCD-like models,
since the heavy
technifermions are providing the bulk of the $W^\pm$ and $Z$ masses.  A large
value of  $T$ can still be a
problem in multiscale models, if equation (\ref{ratio}) holds.  From Table 1,
one can also see that the $N$
has more attractive interactions than the $E$, but that they both have fairly
weak ETC interactions, thus I
expect that the $N$ will be slightly heavier than the $E$.  This is perhaps
surprising.  It suggests that the ETC corrections to the $Z\nu_\tau
\overline{\nu_\tau}$ coupling
could be larger than the ETC correction  to the $Z \tau \overline{\tau}$
coupling.

In order to quantify  how bad isospin splitting for techniquarks can be, I will
resort to a crude
approximation of using a single techniquark-loop (with constant masses) as an
estimate of
$T$.  This is expected to be an overestimate, since technifermion masses should
actually fall with
increasing momentum.  My hope-othesis (as opposed to hypothesis)  is that this
estimate is within a factor of 2 of the correct answer.
The one-techniquark-loop graph gives \cite{Drho}:
\begin{equation}
\Delta\rho_*\equiv \alpha T = {{d \,N_c \,a}\over{16 \pi^2 f^2}}
\left(m_U^2 + m_D^2 - {{4 m_U^2 m_D^2}\over{m_U^2-m_D^2}} \ln\left({m_U\over
m_D}\right)\right)~,
\label{Deltarho}
\end{equation}
where $d$ is the dimension of the technifermion representation, $f=246$ GeV,
$N_c=3$ is the number of
colors, and $1/2 < a < 2$ is the assumed factor of 2 uncertainty.  Now, given a
fixed mass ratio $m_D/m_U$ and an
upper bound on $\alpha T$, we can find an upper bound on $m_U$ (and $m_D$).
For example for
$m_D/m_U = 0.5$ and $\alpha T < 0.01$, we have $m_U < 156 /\sqrt{a}\, {\rm
GeV}< 221 {\rm GeV}$.
Thus there is substantial room for isospin breaking in this type of model,
given that we expect $m_U >
m_t$.  As $m_D/m_U$ approaches unity, the bound, of course, weakens.

The result that $m_U$ is so close to $m_t$ brings to light the
central problem of multiscale models:  how can the $t$ quark be so heavy?  In
order to achieve $m_t \approx 175$ GeV would necessitate some very nasty
fine-tuning of ETC interactions in this toy model.  In vector-like multiscale
models the isospin problem is much worse, even assuming that the correct value
of $m_t$ can be produced.  Using equation
(\ref{ratio}) we expect that $m_D / m_U < 0.3$, which gives a stronger bound:
 $m_U < 114/\sqrt{a} < 161 {\rm GeV}$.  In order to avoid this absurdity,
realistic, vector-like, multiscale models
must have different ETC scales for the $t$ and the $b$, which requires even
more complicated  models.

As for the techniquark contribution to  $S$, using the methods of ref.
\cite{revenge}
I find:
\begin{eqnarray}
S_{TQ}&=&{{-a^\prime \,Y\,N_c \,d }\over{\pi}}\int_0^1 dx
\ln\left({{m_U^2-x(1-x)M_Z^2}
\over{m_D^2-x(1-x)M_Z^2}}\right)x(1-x)\nonumber \\
& &+{{a^\prime\,N_c\, d}\over{2\pi}}{{m_U^2}\over{M_Z^2}}
\int_0^1 dx \ln\left({{m_U^2}
\over{m_U^2-x(1-x)M_Z^2}}\right) \nonumber\\
& &+{{ a^\prime\,N_c\, d}\over{2\pi}}{{m_D^2}\over{M_Z^2}}
\int_0^1 dx \ln\left({{m_D^2}
\over{m_D^2-x(1-x)M_Z^2}}\right)~,
\label{STQ}
\end{eqnarray}
where $a^\prime$ is another factor of 2 uncertainty.  For $m_D = 100$ GeV, and
$m_U = 200$
GeV, $S_{TQ} = 0.35 a^\prime$, which is about half of the heavy, degenerate
techniquark estimate:
$S_{TQ} = a^\prime N_c\, d /(6 \pi)$.

\section{Conclusions}
Chiral TC models take the multiscale paradigm to an extreme.
While the  toy model I have discussed illustrates the possibility of a chiral
TC model,
it is far from satisfying.  The most glaring defect is that isospin breaking
is put in by hand, since there is no $SU(2)_R$ symmetry in the quark
and lepton sector.
One might have hoped for a dynamical explanation.  Of course it is possible
that at some higher scale dynamics could act in a theory with an
$SU(2)_R$ symmetry and produce an effective theory where part of the
$SU(2)_R$ doublet is essentially replaced by some other fermions in a
different ETC representation.  This is what is supposed to happen for
neutrinos in the model of ref. \cite{AppelTern}.

On the positive side,
chiral TC models offer a simple
way to split the $t$ and $b$ quarks without fine-tuning, while at the same time
reducing the required splitting between the $U$ and $D$ technifermions.  In
principle
the ETC contribution to $\Gamma(Z\rightarrow b \overline{b}) $ can be reduced
by
(up to) a factor of 4.  The techniquark contribution to the $S$ parameter can
also be
reduced. Since the phenomenology of chiral TC models
is almost entirely unexplored, there may be more interesting chiral TC models,
and possibly
one that actually works.

\vskip 0.15 truein

\noindent\medskip{\bf Appendix on the inadequacy of the MAC analysis}

While discussing the $SU(4)_{TC}$ gauge theory with the fermion content
listed in lines (\ref{ferm41L}) through (\ref{ferm43}), I
assumed the MAC analysis did not yield the correct fermion condensate.
 The MAC is ${\bf \overline{10}} \times {\bf 4} \rightarrow {\bf
\overline{4}}$.
This condensate would break
$SU(4)_{TC}$ down to $SU(3)$.  I assumed that the condensation occurs in the
channel ${\bf 4} \times {\bf 4} \rightarrow {\bf 6}$, which breaks\footnote{A
condensate
in the ${\bf 6}$ could also break the gauge group down to $SU(2) \otimes
SU(2)$.  If the gauge
coupling were weak, then vacuum alignment \cite{vac} would prefer  $Sp(4)$,
since it is a larger
group and keeps more $SU(4)$ gauge bosons massless.  I am assuming  that
$Sp(4)$ is also
preferred in the strong coupling problem at hand.}
$SU(4)_{TC}$ down to $Sp(4)_{RTC} $.   (I will leave consideration of the
channel
${\bf 4} \times {\bf \overline{4}}  \rightarrow {\bf 1}$ until the end of the
appendix.)
The first point is that while the
assumed condensation channel has a smaller $\Delta C_2$, it leaves a larger
gauge
symmetry than the MAC does.  Perturbatively,
generating masses for gauge bosons increases rather than decreases the
vacuum energy.  Furthermore, the MAC analysis may be going astray here
since it counts all gauge bosons equally, whether the condensate gives
them a mass or not.  If we only count contributions from massless gauge
bosons we find  $\Delta C_2\, |_{\rm massless}({\bf \overline{10}} \times {\bf
4}
\rightarrow {\bf \overline{4}})  = 8/3$, while $\Delta C_2 |_{\rm massless}
({\bf 4} \times {\bf 4} \rightarrow {\bf 6}) = 5/2$, which is
almost as strong .  However, there is an additional complication, which is
probably the most important consideration.  The channel
${\bf 4} \times {\bf 4} \rightarrow {\bf 6}$ will give masses to 10
Dirac fermions while the MAC
gives masses to only one Dirac fermion.  While the dynamical mass produced by
the ${\bf \overline{10}} \times {\bf 4} \rightarrow {\bf \overline{4}}$
channel will be somewhat larger than that produced by the ${\bf 4} \times
{\bf 4} \rightarrow {\bf 6}$, the contribution to the effective potential
\cite{CJT}
will be far outweighed by the sheer number of fermions contributing in
the later channel!

At first sight, one might expect that the channel
${\bf 4} \times {\bf \overline{4}} \rightarrow {\bf 1}$ would condense at a
slightly
higher scale than
${\bf 4} \times {\bf 4} \rightarrow {\bf 6}$. If it did, it would not affect
the
arguments above regarding the pattern of gauge symmetry breaking.  However we
are dealing with strong gauge couplings here, so it is quite plausible that the
broken
TC gauge boson masses are larger than the mass scale associated with
condensation.
In particular, the naive critical coupling in the assumed channel is
\begin{equation}
\alpha_c({\bf 4} \times {\bf 4} \rightarrow {\bf 6}) = {{2 \pi}\over
{3 \Delta C_2}} = {{8 \pi} \over {15}}~,
\end{equation}
which corresponds to a gauge coupling $g \approx 4.5$.  Thus the broken TC
gauge
bosons should have a mass larger than the fermions which condense in the
${\bf 4} \times {\bf \overline{4}} \rightarrow {\bf 1}$ channel.  Whether one
says this
channel condenses before or after TC breaks is largely a matter of definition.

\noindent\medskip\centerline{\bf Acknowledgments} \vskip 0.15 truein I
thank  S. Chivukula, M. Dugan, K. Lane, R. Sundrum and L.C.R. Wijewardhana for
helpful discussions.
This work was partially supported by the Department of Energy under contracts
\#DE-AC02ERU3075, and \#DE-FG02-91ER40676, and by the National Science
Foundation under grant \#PHY89-04035.  I am indebted to
the Institute for Theoretical Physics, Santa Barbara, where part of this work
was
completed during the ``Weak Interactions" workshop.

\end{document}